\documentclass[overcite,leqno,fleqn,titlepage,11pt]{article}

\usepackage[dvips]{graphicx}
\usepackage{floatflt}
\usepackage{epsf}
\usepackage{fancyheadings}
\usepackage{overcite}

\rmfamily

\begin{document}
\begin{center}
%

\LARGE
\textbf{Broad pore lifetime distributions: \\ A fundamental concept for cell electroporation} 
\normalsize

\vspace{0.09in} 

Julie V. Stern$^1$,
Thiruvallur R. Gowrishankar$^1$,
Kyle C. Smith$^{1}$, 
and
James C. Weaver$^{1,*}$,

$^1$Harvard-MIT Division of Health Sciences and Technology,
Institute for Medical Engineering and Science,
Massachusetts Institute of Technology, Cambridge, MA, USA;
$^*$Corresponding author

\vspace{0.12in}
\end{center}

\vspace{0.23in} 
\Large
\textbf{Abstract} 
\normalsize

\textbf{We describe a concept that has the potential to change how we think about the underlying mechanisms of cell membrane electroporation (EP).
Prior experimental, theoretical and modeling have emphasized a single pore lifetime as adequate for particular conditions.
Here we introduce a much more complex response:  The rapid creation of many types of pore structures, of which some are traditional transient lipidic pores (TPs),
but the great majority are complex pores (CPs) based on both lipids and other molecules or molecular segments.
At the inner leaflet of the cell plasma membrane (PM) non-lipidic molecules come from the over-crowded cytoplasm.
At the outer leaflet they originate from the extracellular medium and extracellular matrix.
Some partially or fully insert into TPs during or shortly after TP formation, or bind to the membrane nearby.
This process is complex, leading to mostly short-lived structures, with relatively few lasting for long times.
We speculate that the characteristic pore lifetimes range from
$\mathbf{\sim}$100~ns to 1,000~s, based on implications from experiments.
The frequency-of-occurrence probably falls off extremely rapidly with increasing lifetime,
$\mathbf{\tau_{\mathbf{CP}}}$,
which implies that most are inaccessible to traditional experimental methods.
It also suggests that unexpected behavior can occur early in pulsing, 
vanishing before post-pulse observations begin.
}

\vspace{0.23in} 
\Large
\textbf{Introduction} 
\normalsize

\large
\textbf{Distinguishing cell electroporation (EP) from EP in other membrane systems} 
\normalsize

We present a hypothesis that involves greatly increased complexity of pore creation, evolution and eventual destruction.
We believe this is necessary to explain experimental results for cell electroporation (EP), which we explicitly distinguish from vesicular and artificial planar bilayer membrane EP.
%
%
Although investigation and application of cell EP has been pursued for decades
\cite{NeumannSowersJordan_Book_ElectroporationElectrofusionInCellBiology_Plenum1989,%
ChangEtAlEporeBook1992,%
PakhomovMiklavcicMarkov_AdvancedElectroporationTechniquesBiologyMedicine_CRC2010},
it is generally accepted that the basic mechanisms are poorly understood.
For perspective we briefly describe EP generally, and then focus on our hypothesis that cell EP should be distinguished from EP in simpler systems.

It is generally agreed that in all membrane systems there are three stages:
(1) pore creation during pulsing, with a highly non-linear dependence on transmembrane voltage,
$\mathrm{\Delta \phi_{\mathrm{m}}}$,
(2) pore expansion/contraction, also affected by
$\mathrm{\Delta \phi_{\mathrm{m}}}$, and
(3) pore persistence and eventual destruction, usually considered for
$\mathrm{\Delta \phi_{\mathrm{m}}}$ $=$ 0 (full depolarization), with post-pulse membrane resealing attributed to pore lifetimes.
Of these pore creation is reasonably well understood, including agreement between molecular dynamics (MD) and continuum models for
transmembrane voltages and associated membrane fields where these very different methods overlap
\cite{VasilkoskiEtAl_ElectroporationAbsouteRateEquationNanosecondPoreCreation_PhysRevE2006}.
Behavior during pulsing is only occasionally measured
\cite{KinositaEtAl_ElectroporationVisualizedPulseLaserFluorescenceMicroscope_BPJ1988,%
Frey_PlasmaMembraneVoltageChangesDuringNanosecondPulsedElectricFields_BPJ2006,%
FlickingerEtAlFrey_PlantCellsProtoplastsVoltageSensitiveDyeMembraneVoltages_Protoplasma2010}
due to significant technical difficulties, and has therefore received little attention.
%

%
\vspace{0.12in} 
\large
\textbf{Pore destruction and associated lifetimes} 
\normalsize

We emphasize pore lifetime, a basic feature of pore destruction. 
It  is usually considered as the mean value of an exponentially decaying membrane conductance or permeability.
A short perspective
\cite{draftPoreLifetimesShortDiscussionJUL2017_Bchem2017}
makes the case that the traditional view of pure lipidic pores vanishing stochastically with a long lifetime is not correct.
The basic argument is that molecular dynamics (MD) publications show that a basic assumption of the traditional model is invalid,
and in addition shows many examples of lipidic pores vanishing with a $\sim$100~ns lifetime.
This has a major implication.
For example, if there are $\sim 10^6$~pores during a pulse, within about $2 \mathrm{\, \mu s}$ after pulsing
the probable number of pores is $\sim$0.001 (essentially zero)
\cite{draftPoreLifetimesShortDiscussionJUL2017_Bchem2017}.
%
Important MD examples include an illustrative progression of pore structures leading to pore destruction
\cite{LevineVernier_LifeCyclePoreStepsCreationAnnihilation_JMemBiol2010},
an associated pore energy landscape as a function of pore radius that shows a negligible barrier at all to pore destruction
\cite{WohlertEtAl_FreeEnergyPoreLinearTermMD_JChemPhys2006}
and an exponential decay fit to a series of pore closures (destructions) that supports a
$\sim$100~ns lifetime
\cite{BennettEtAlTieleman_AtomisticSimulationsPoreFormationClosure_BLM_BPJ2014}.

\vspace{0.12in} 
\large
\textbf{The concept of complex cell membrane pores} 
\normalsize

In Figure~1 we show an initial, incomplete picture of a broad lifetime distribution that we propose for cell membrane pores created by large physical perturbations.
Although we are mainly concerned with applied electric fields, mechanical perturbations that increase membrane tension may also be involved
\cite{Evans_NewMembraneConceptAppliedToFluidShear-LyticTension_BPJ1973,%
EvansSmith_KineticsHoleNucleation-BiomembraneRupture_NewJPhys2011}.
%
%
We envision involvement of both purely lipidic pores (TPs), the traditional view, and also complex pores (CPs) that involve other types of molecules.
Presently we do not know what CP structures emerge, as there is a huge, ``biologically large'', number of
combinations of biological molecules in various states, configurations and binding strengths.
\cite{Frauenfelder_BiologicallyLargeNumbers-GreaterThanAstronomicallyLargeNumbers_Proteins-ParadigmsOfComplexity_PNAS2002}.
For very large electric field strengths a collection of unusual membrane openings and other structures has been explicitly suggested
\cite{PliquettEtAl_HighElectricFieldEMicellsCellMembranes_BChem2007}.
Overall, the response of the plasma membrane (PM) to two complex environments, each with a large number of molecules, should be
markedly different than a pure lipid membrane contacting only aqueous electrolyte-based media, even if some have full or residual content from cell growth media.

The most complex, and realistic, condition is that of the PM in a living tissue.
The inner leaflet is in direct contact with the overcrowded cytoplasm, and
the outer leaflet is intimately exposed to the extracellular medium and also contacts the extracellular matrix.
It is known that the cortex (cytoskeleton) interacts with the PM during mitosis,
and that the cytoskeleton is suggested to limit expanded pore size
\cite{KrassenPliquettNeumann_SingleCHOcellNonlinearIVcurveElectroporation60nmPore_BChem2007}.
These and other structures near and or contacting cell membranes provide our general motivation to consider CPs with diverse behavior
even if we know little about their structures.

\vspace{0.12in} 
%
\begin{figure}[!h]
\begin{center}
\begin{tabular}{l}
\includegraphics[width=6.00in]{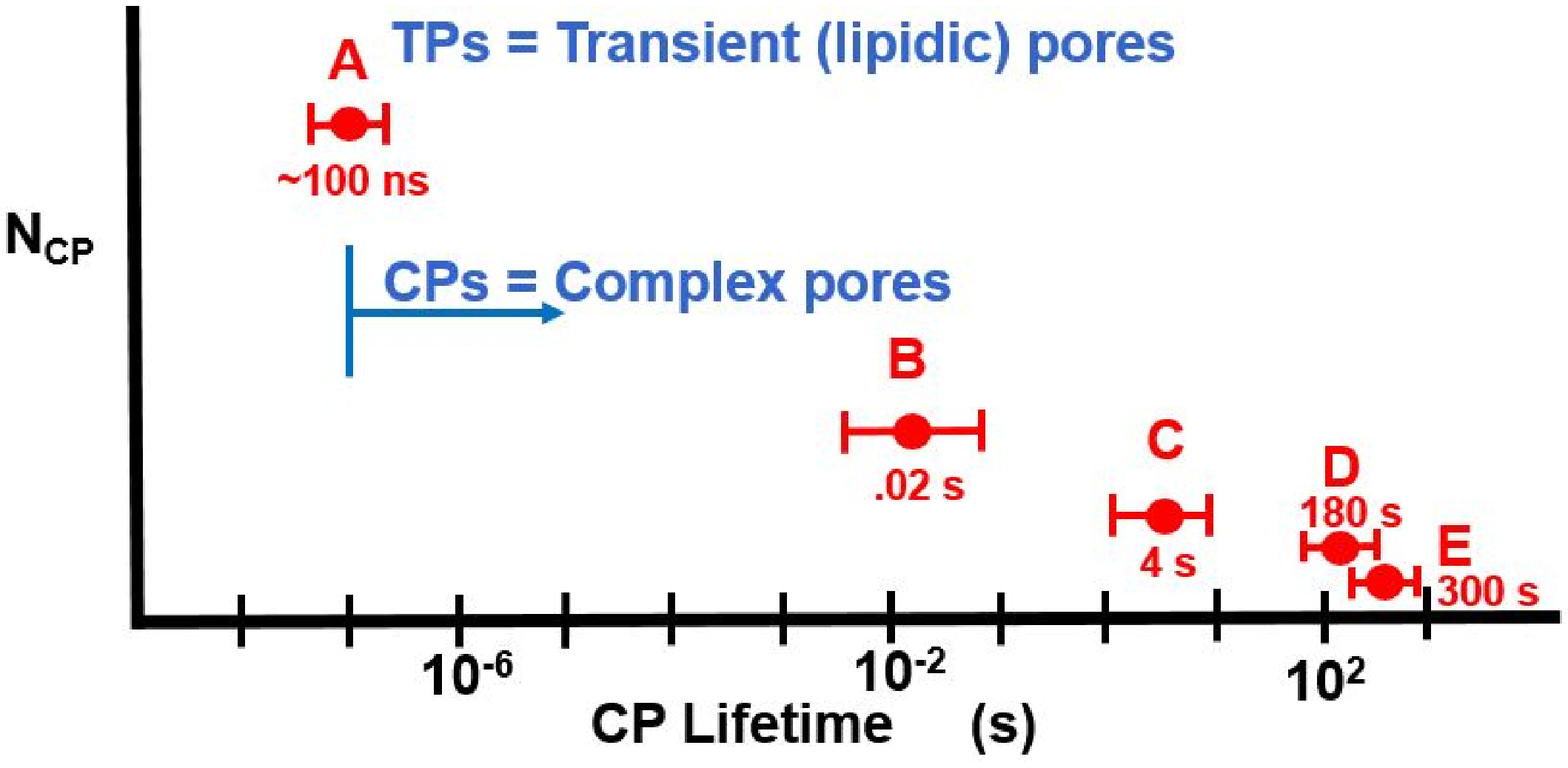}\\
\end{tabular}
\end{center}
\small
%
%
\vspace*{-0.22in} 
\textbf{Figure~1: Broad lifetime distribution for CPs and TPs, a new fundamental concept.} 
Electroporation (EP) experiments typically report post-pulse membrane recovery by a single relaxation quantity.
Single resealing times are therefore common, often translated into particular pore lifetimes.
Here we challenge this traditional view, hypothesizing that from the early times when pore creation occurs over a finite, but often short interval,
there is inherently a wide distribution of permeabilizing structures with a broad distribution of lifetimes.
Based on the experiments cited below we expect lifetimes ranging from $\sim$100~ns to  $\sim$1,000~s, an impressive 10 orders of magnitude.
Some complex events may occur as TPs and CPs are simultaneously formed, with molecules and molecular segments drawn into CPs as they form.
Such structures may be fleeting, but may contribute to solute rectification, with net transport persisting long afterwards.
We argue
\cite{draftPoreLifetimesShortDiscussionJUL2017_Bchem2017}
that traditional lipidic pores have lifetimes,
$\tau_{\mathrm{p}}$, of order $100$~ns.
Further, some CPs may form simultaneously with TPs (red \textbf{A}; with rough error bars).
A longer
$\tau_{\mathrm{p}}$ lifetime estimate is 0.02~s \textbf{B},
based on recovery between pulses in a pulse train
\cite{PakhomovEtAlPakhomova_MultipleNanosecondPulsesIncreaseNumber-NotLongLivedPoreSize_BBA2015}.
Very different experiments lead to a lifetime of $\sim$4~s \textbf{C}
\cite{SmithKC_CellModelWithDyamicPoresAndElectrodiffusionofChargedSpecies_DoctorateThesisMIT_2011},
based on modeling of calcein uptake by prostate cancer cells
\cite{CanatellaEtAlPrausnitz_ElectroporationUptakeViabilityQuantitativeStudy_BPJ2001}.
Another important experiment
\cite{PakhomovEtAl_nsPEF_LongLastingSinglePulsePermeabilizationPM_BEMS2007},
reports that for its particular conditions the relaxation time for PM recovery corresponds to
$\tau_{\mathrm{p}}$ $\approx$ 180~s \textbf{D}.
Two experiments
\cite{KennedyEtAlBooske_QuantificationElectroporationKineticsPI_FinalVersion_BPJ2008,%
PakhomovaEtAlPakhomov_Electroporationn-InducedElectrosensitization_PLoS_ONE2011}
show a remarkable delayed increase in propidium uptake at very long times,
$\sim 300 \pm \sim 300$~s, \textbf{E}.
Both the delays and their variability suggest that only a few CPs survive.
For a cell-pulse combination that creates $10^5$ to $10^7$~pores with $\sim$100~ns lifetimes, $\tau_{\mathrm{p}}$, all vanish in $\sim$2~ns.
We also estimate that only a few pores (10 to 100) are present at $\sim$1,000~s.
More experiments that allow lifetime estimates are needed to determine whether $\tau_{\mathrm{p}}$ values follow a common curve,
which is likely to fall rapidly for small  $\tau_{\mathrm{p}}$ and slows its decrease for large  $\tau_{\mathrm{p}}$.
All lifetimes are presumed present during pulsing that creates a short burst of pores,
with short lifetime CPs quickly vanishing, with the distribution loosing CPs at a slowing rate according to their $\tau_{\mathrm{p}}$ values.
There is some similarity to nuclear explosions in an environment that creates a large number of different radioactive isotopes,
and many of these isotopes have a very broad range of lifetimes.
Some of the corresponding literature may be relevant to the present problem.
A large number of CPs with the shorter lifetimes may control transport briefly, by moving a solute just half-way across the membrane. 
If CP creation involves bringing a transported molecule into the CP during its formation,
upon CP closure the molecule has been transported.
\end{figure}
\vspace{0.12in} 

%
Although we argue that traditional TPs vanish within
$\sim 2 \mathrm{\, \mu s}$ after a well defined pulse
\cite{draftPoreLifetimesShortDiscussionJUL2017_Bchem2017},
we extend our thinking to the possibility for cell membranes close to many molecules of different size, shape (including branching) and charge.
This leads to the expectation than large local fields associated with pulsing are heterogeneous on a nanoscale, and these
together with fluctuations can lead to a wide variety of membrane openings, complex pores (CPs).
We suggest that CPs can contribute to post-pulse behavior for times well beyond the $2 \mathrm{\, \mu s}$ estimate for TPs.

\large
\textbf{CPs are ``accidental'' pores, in analogy with accidental cell death} 
\normalsize 

\vspace{0.12in}
Further, CPs are ``accidental'', with presently unknown structures, perhaps rapidly changing, due to random encounters due to the large physical perturbation.
The CP structures are not the result of evolution, but are accidental in the same sense that accidental cell death is not regulated,
but due instead to large physical interactions
\cite{GalluzziEtAlKroemer_Essential-vs-AccessoryAspectsCellDeath_Recommendations-NCCD2015_SHORT_CellDeathDiff2015}.
This unregulated process is likely to lead to a wide distribution in binding or association energies, with most weak, corresponding to short lifetimes.
A consequence is an extremely broad range of CP lifetimes (Figure~1).
%

%
\vspace{0.12in} 
\large
\textbf{Experiments that suggest CPs exist}
\normalsize

Importantly, there is experimental evidence that CPs can exist for varying times.
These are characteristic times 
with little information regarding how many pores are involved.
The bold upper-case letter \textbf{A - E} in the caption of Figure~1 are associated with experiments, discussed below.

Several experiments support the orders of magnitude in \textbf{B} through \textbf{E}.
There are additional experiments which report evidence for pores that cannot be traditional TPs 
\cite{draftPoreLifetimesShortDiscussionJUL2017_Bchem2017}, and also some conjecture for CP existence.
This diverse evidence is presented chronologically.

%
%
In 2007 an important experiment described the post-pulse recovery of the PM,
reporting that the conductance returned to (within experimental noise) its initial value in $\sim 15 \mathrm{\, m }$ (900~s)
\cite{PakhomovEtAl_nsPEF_LongLastingSinglePulsePermeabilizationPM_BEMS2007}.
We conservatively assume that 5 exponential lifetimes corresponds to essentially full recovery,
and this yields a recovery (resealing) lifetime of $\sim 180 \mathrm{\, s}$.
This is a very long lifetime compared to MD.

%
%
In 2008 rather different experiments
\cite{KennedyEtAlBooske_QuantificationElectroporationKineticsPI_FinalVersion_BPJ2008}
reported delayed increases of single cell PM permeability following a single pulse for $40 \mathrm{\, \mu s}$ for various electric field strengths.
The permeability increases were inferred, based on quantified propidium uptake,
at post-pulse times of order $\sim 300 \pm \sim 300$~s, a huge range in which measurement times extended to 1,800~s.
Additional results by the same group were included in a later publication
\cite{KennedyEtAlMurphy_CationicPeptideEnhancesElectroporation-IRE-Motivation_PLoS-ONE2014},
with and without peptide-altered behavior.

%
In 2009 experiments were reported
\cite{PakhomovEtAl_LongLivedLipidNanoporesIonChannel-Like_BBRC2009}
with the interpretation that ``...nanopores can form a stable, ion channel-like conduction
pathway in [a] cell membrane'' by nsPEF pulsing protocols.
%
This is an explicit claim that CPs are created, not traditional lipidic pores.

In 2011 additional experiments
\cite{PakhomovEtAl_LongLivedLipidNanoporesIonChannel-Like_BBRC2009}
reported delayed uptake of propidium, but in this case the perturbation/stimulus was a pulse train with strengths of 1.8 to 12.3~kV/cm,
durations of 60~ns to $9 \mathrm{\,\mu s}$, and 2 to 3,750 pulses applied sequentially.
Strikingly, these experiments report delayed permeability increases similar to those found earlier for single, $40 \mathrm{\, \mu s}$ pulses
\cite{KennedyEtAlBooske_QuantificationElectroporationKineticsPI_FinalVersion_BPJ2008}.

%
In 2015 an experimental study reported significant recovery between the individual pulses of a pulse train.
This provides evidence that some long-lived pores are involved,
with their longevity related to the capability to create more pores after the inter-pulse interval of $\sim 10^{-2} \mathrm{\, s }$ new pores are created
\cite{PakhomovEtAlPakhomova_MultipleNanosecondPulsesIncreaseNumber-NotLongLivedPoreSize_BBA2015}.
The concept of evaluating experiments in terms of more pores created after a gap between pulses may provide a powerful approach to quantitative examination of CP lifetimes,
perhaps limited mainly by technical issues such as the constraints on separations of the individual members of a pulse train,
and the likelihood that between the individual pulses 
$\Delta \phi_{\mathrm{m}}$ is not zero (full depolarization is not realized). 
%
%
We have reported a modeling approach for assessing the degree of recovery between the individual members of a pulse train
\cite{SonEtAl_PulseTrainConventionalEP-20FoldDecreaseThenSawtooth-CalciumUpake-Electrosensitization_TBME2015},
which should be applicable to this proposed approach for determining CP lifetime ranges at particular times after pulsing.
Overall, the use of pulse trains with a wide range of intra-pulse spacings may provide a powerful approach to determining part of a broad lifetime distribution,
at least approximately.

Recently a rather different type of experiment reported that post-pulse contains a contribution from active transport, not only diffusion
\cite{SozerLevineVernier_QuantitativeLimits_SmallMoleculeTransport-Electropermeome_MeasuringModeling-SingleNanosecondPerturbations_SciRep2017}.
The conventional view is that post-pulse the PM is fully depolarized, so that post-pulse transport is thereafter purely passive.
The 2017 report shows otherwise.
It also provides support for one or more types of long-lived permeability states that exist long after TPs have vanished.

\large
\vspace{0.12in} 
\textbf{Summary}
\normalsize

Single lifetimes reported from experiments can be long, but reporting the single value distracts the reader from the possibility that there may be a lifetime distribution.
The important distinction is that the broad lifetime distribution (BLD) hypothesis allows early behavior by numerous short lifetime CPs to
contribute to behavior (e.g. solute transport that persists), but these CPs are not observed because of their fleeting existence.
They can, however, possibly contribute to molecular transport, leaving evidence of their existence and their potential importance to understanding cell electroporation
by determining the molecular and ionic transport these CPs allowed.

\vspace{0.23in} 
\Large
\textbf{Acknowledgments} 
\normalsize

This work was supported by AFOSR MURI grant FA9550-15-1-0517 on Nanoelectropulse-Induced
Electromechanical Signaling and Control of Biological Systems, administered through Old Dominion University.

We thank
P. T. Vernier,
E. S\"{o}zer,
R. S. Son
and
A. E. Esser
for stimulating discussions,
and K. G. Weaver for computer support.

%
\vspace{0.32in} 
\textbf{References}
\vspace*{-0.52in} 
\small
\def\refname{}
\bibliographystyle{unsrt}

\normalsize

\end{document}